Research Article

Yassir El Moutaoukal, Rosario R. Riso, Andrea Bianchi, and Henrik Koch*

# Unveiling chiral electron–photon correlation effects in circularly polarized optical devices

**Abstract:** Strong coupling with circularly polarized vacuum fluctuations offers a viable route to manipulate molecular chirality. While experiments are advancing toward the realization of chiral cavities, a mean-field theoretical framework for describing electron–photon interaction in this platform has been missing. Here, we present a mean-field theory that can be systematically improved to capture the chiral correlation effects responsible for the enantioselective power of chiral light. We use strong coupling Møller-Plesset perturbation theory for accessing the excitation manifold of electrons and chiral virtual photons. We apply the developed methods to selected chiral systems and show that the mean-field theory captures cavity frequency dispersion, but fails to describe the chiral discrimination arising from coupled electron–photon excitations.

## 1 Introduction

Strong coupling between light and matter represents a promising route for modifying molecular properties in a non-invasive fashion [1–5]. The strong coupling regime arises when the interaction between matter and the electromagnetic field is strong enough to entangle them, overcoming dissipation and decoherence processes [6, 7]. In this regime, the quantum nature of the fields must be explicitly accounted for, necessitating a quantum electrodynamics (QED) treatment. In particular, quantum vacuum fluctuations, or virtual photons, can dress molecular excitations, giving rise to hybrid light–matter states known as molecular polaritons [8]. These quasiparticles exhibit properties that go beyond those of the individual constituents, and their characteristics can be finely tuned by adjusting the electromagnetic field [9–11]. To model the molecular polaritons, *ab initio* quantum chemistry methods have recently been extended to include QED environments [12–21]. An especially intriguing scenario emerges when vacuum photons are circularly polarized. Such light can discriminate between enantiomers, the two non-superimposable mirror images of a chiral molecule. In this way, the symmetry that matter typically exhibits is broken. In the electronic strong coupling regime, where the frequency of the chiral photons lies in the UV–visible range, this symmetry breaking can lead to the formation of novel ground and excited state energy surfaces, opening previously unexplored pathways for polaritonic control of enantioselective reactivity [22–25]. In line with this prospect, recent studies have also demonstrated that a single monochromatic circularly polarized attosecond laser pulse can induce long-lived electronic chiral currents in oriented achiral molecules [26–28], paving the way to the possibility of leaving a detectable imprint of the coupling with chiral virtual photons even in achiral molecules, where such effects would otherwise remain hidden.

Experimentally, strong coupling with chiral vacuum fluctuations has been pursued in a quest to design optical devices that can confine only one circular polarization of the electromagnetic field within a small quantization volume [29–33]. To this end, researchers have developed mirrors incorporating chiral metamaterials, which exhibit structure-dependent differential reflectance for circularly polarized light over a broad range of wavelengths [34–36]. Alternative strategies include embedding thin layers of chiral materials into conventional Fabry–Pérot cavities [37]. Despite these advances, the reliable fabrication of fully functional chiral cavities remains a significant challenge, underscoring the importance of theoretical studies to guide future developments [38–40]. So far, the theoretical modeling of these complex many-body systems where electrons and chiral virtual photons interact, has re-

**Yassir El Moutaoukal,** Department of Chemistry, Norwegian University of Science and Technology, 7491 Trondheim, Norway
**Rosario R. Riso,** Department of Chemistry, Norwegian University of Science and Technology, 7491 Trondheim, Norway
**Andrea Bianchi,** Scuola Normale Superiore, Piazza dei Cavalieri, 7, 56126 Pisa PI, Italy
**\*Corresponding author: Henrik Koch,** Department of Chemistry, Norwegian University of Science and Technology, 7491 Trondheim, Norway, e-mail: henrik.koch@ntnu.no

lied either on simplified model Hamiltonians or on correlated *ab initio* approaches [23, 41–46]. However, no mean-field wave function method has yet been introduced to describe such systems, and a comprehensive analysis of the electron–photon correlation responsible for the enantioselective power of chiral light remains to be developed.

In this work, we address this gap by extending the strong coupling QED Hartree Fock wave function parameterization to a Hamiltonian tailored for modeling the enantiomeric discrimination power inside chiral cavities. This development establishes a mean-field framework that can be systematically improved using strong coupling Møller-Plesset perturbation theory, enabling a more detailed analysis of the correlation effects between electrons and chiral virtual photons. The paper is organized as follows: in Section 2 we develop the theoretical framework, followed in Section 3 by details of the implemented methodology; in Section 4 we present and discuss the results, finally giving our conclusions and future perspectives in Section 5.

## 2 Theoretical modeling

In this section, we develop the theoretical framework for modeling chiral molecular systems interacting with circularly polarized virtual photons within chiral cavities. We begin by deriving the light–matter Hamiltonian from the minimal coupling formulation of QED. Then, we introduce the wave function parametrization used to describe the ground state properties of the interacting system. Finally, we present a Møller–Plesset perturbative theory approach to systematically incorporate electron–electron, chiral electron-photon and photon-photon correlations beyond the mean-field description.

### 2.1 Hamiltonian for chiral cavities

The interaction between light and matter is modeled using the Pauli-Fierz molecular Hamiltonian in the Born-Oppenheimer approximation [47]

$$H = \frac{1}{2}\sum_j (\mathbf{p}_j - \mathbf{A}(\mathbf{r}_j))^2 + \frac{1}{8\pi}\int (\mathbf{E}^2(\mathbf{r}) + c^2\mathbf{B}^2(\mathbf{r})]) d^3r$$
$$+ \sum_{j>k} \frac{1}{|\mathbf{r}_j - \mathbf{r}_k|} + \sum_{I>J} \frac{Z_I Z_J}{|\mathbf{R}_I - \mathbf{R}_J|} - \sum_{jI} \frac{Z_I}{|\mathbf{r}_j - \mathbf{R}_I|}, \quad (1)$$

where the electronic momenta $\mathbf{p} = -i\boldsymbol{\nabla}$ are minimally coupled to the vector potential $\mathbf{A}(\mathbf{r})$ generating the electric and magnetic fields $\mathbf{E}(\mathbf{r})$ and $\mathbf{B}(\mathbf{r})$. In order to model cavity effects on the electronic degrees of freedom, it is important to treat the electromagnetic field within QED theory [47, 48]. The canonical quantization of photons gives

$$\frac{1}{8\pi}\int [\mathbf{E}^2(\mathbf{r}) + c^2\mathbf{B}^2(\mathbf{r})] d^3r \xrightarrow{\text{QED}} \sum_{\mathbf{k},\lambda} \omega_{\mathbf{k}} b^\dagger_{\mathbf{k},\lambda} b_{\mathbf{k},\lambda}, \quad (2)$$

where $b^\dagger_{\mathbf{k},\lambda}$ and $b_{\mathbf{k},\lambda}$ create and annihilate a photon for the $(\mathbf{k},\lambda)$ mode of the electromagnetic spectrum. Here $\mathbf{k}$ is the wave vector, $\lambda$ labels the two possible transverse polarization states, and $\omega_{\mathbf{k}}$ is the frequency of the mode. The vector potential is promoted to an operator where the Fourier expansion within a quantization volume $V$ is given

$$\mathbf{A}(\mathbf{r}) = \sum_{\mathbf{k},\lambda} \sqrt{\frac{2\pi}{V\omega_{\mathbf{k}}}} \left[ \boldsymbol{\epsilon}_{\mathbf{k},\lambda} b_{\mathbf{k},\lambda} e^{i\mathbf{k}\mathbf{r}} + \boldsymbol{\epsilon}^*_{\mathbf{k},\lambda} b^\dagger_{\mathbf{k},\lambda} e^{-i\mathbf{k}\mathbf{r}} \right], \quad (3)$$

where $\boldsymbol{\epsilon}_{\mathbf{k},\lambda}$ is the field polarization for the $(\mathbf{k},\lambda)$ mode. We stress that the commonly employed dipole approximation is insufficient to describe the enantioselective power of chiral light, as terms up to first order in the expansion are required to account for the contribution from magnetic moment to the interaction. However, Taylor expansion of the exponentials introduces gauge-invariance issues and we therefore consider the full spatial shape of the field. In order to specify the Hamiltonian for a chiral cavity of volume $V$, formed with mirrors acting as perfect reflectors, we restrict the expansion to include only two counter-propagating modes in the $\hat{z}$ direction, $\mathbf{k}$ and $-\mathbf{k}$, whith the same frequency $\omega$ and opposite handedness of the chiral polarization vector

$$\boldsymbol{\epsilon}_{\pm} = \frac{1}{\sqrt{2}}\begin{pmatrix} 1 \\ \pm i \\ 0 \end{pmatrix}. \quad (4)$$

In this way, the molecular system is embedded in the electromagnetic field of a chiral standing wave with vector potential

$$\mathbf{A}(\mathbf{r}) = \frac{\lambda}{\sqrt{2\omega}}\left(\boldsymbol{\epsilon}_{\pm}(b_{\mathbf{k}} + b^\dagger_{-\mathbf{k}})e^{i\mathbf{k}\mathbf{r}} + \boldsymbol{\epsilon}^*_{\pm}(b^\dagger_{\mathbf{k}} + b_{-\mathbf{k}})e^{-i\mathbf{k}\mathbf{r}}\right) \quad (5)$$

where we introduced the light-matter coupling strength

$$\lambda = \sqrt{\frac{4\pi}{V}}. \quad (6)$$

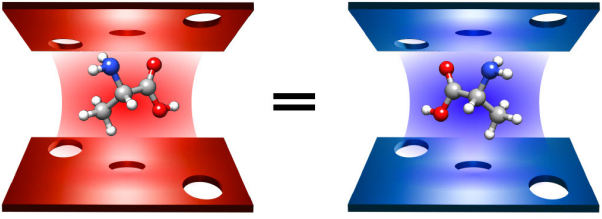

**Fig. 1:** Pictorial representation of right- and left-handed circularly polarized chiral cavities, respectively coupled to the S and R enantiomers of an alanine molecule. The light–matter systems are related by inversion symmetry of the overall chirality and are energetically equivalent in the formulation of Eq. 1.

When the entire light-matter system is reflected, the molecule becomes its mirror image while the cavity polarization reverses, see Figure 1. The system energy is unchanged, since the Hamiltonian contains no parity-violating interaction. Expanding the first term in Eq. 1, we obtain the diamagnetic term $\mathbf{A}^2$ and its inclusion or omission has been widely discussed in the literature, particularly in relation to the emergence of superradiant phases [49–51]. A more critical issue, however, has been raised in the field of cavity QED for extended matter systems. Rokaj *et al.* [52] have demonstrated, through a *no-go* argument, that inside a cavity the inclusion of this term is fundamental to guarantee the existence of a ground state. The same reasoning applies to the length gauge Hamiltonian in the dipole approximation when the dipole self-energy (DSE) contribution is neglected [53, 54]. For one pair of counter-propagating modes, the $\mathbf{A}^2$ term can be reabsorbed with a Bogoljubov transformation

$$\begin{pmatrix} \alpha \\ \beta \end{pmatrix} = \begin{pmatrix} \cosh\theta & \sinh\theta \\ \sinh\theta & \cosh\theta \end{pmatrix} \begin{pmatrix} b^\dagger_{-\mathbf{k}} \\ b_{\mathbf{k}} \end{pmatrix} \quad (7)$$

that introduces two new photonic operators, $\alpha$ and $\beta$, which satisfy the canonical commutation relations

$$[\alpha, \alpha^\dagger] = [\beta, \beta^\dagger] = 1, \quad [\alpha, \beta] = 0, \quad [\alpha, \beta^\dagger] = 0. \quad (8)$$

The angle $\theta$ is given by

$$\tanh 2\theta = \frac{N_{el}\lambda^2}{N_{el}\lambda^2 + 2\omega^2}, \quad (9)$$

where $N_{el}$ is the number of electrons. The effect of this transformation is to introduce a non-size-extensive cavity-dressed frequency

$$\bar{\omega} = \sqrt{\omega^2 + N_{el}\lambda^2}. \quad (10)$$

Moreover, note that this frequency is bounded from below at $\omega = 0$, preventing numerical energy divergencies when approaching the limit. Introducing the operators in Eq. 8, the light-matter Hamiltonian becomes

$$\begin{aligned} H = H_{el} &+ \bar{\omega}(\alpha^\dagger\alpha + \beta^\dagger\beta) \\ &+ \frac{\lambda}{\sqrt{2\bar{\omega}}} \sum_j (\mathbf{p}_j \cdot \boldsymbol{\epsilon}_\pm)e^{i\mathbf{k}\mathbf{r}_j}(\alpha + \beta^\dagger) \\ &+ \frac{\lambda}{\sqrt{2\bar{\omega}}} \sum_j (\mathbf{p}_j \cdot \boldsymbol{\epsilon}_\pm^*)e^{-i\mathbf{k}\mathbf{r}_j}(\alpha^\dagger + \beta). \end{aligned} \quad (11)$$

Introducing second-quantization for the electrons, the electronic Hamiltonian reads

$$H_{el} = \sum_{pq} h_{pq}E_{pq} + \frac{1}{2}\sum_{pqrs} g_{pqrs}e_{pqrs} \quad (12)$$

where we have employed an orbital basis set $\{\phi_p\}$ [55]. In Eq. 12, the one and two electron integrals are given by

$$h_{pq} = \int \phi_p^*(\mathbf{r})\left(-\frac{\boldsymbol{\nabla}^2}{2} - \sum_I \frac{Z_I}{|\mathbf{r} - \mathbf{R}_I|}\right)\phi_q(\mathbf{r})\, d^3\mathbf{r} \\ + \delta_{pq}\frac{h_{nuc}}{N_e}, \quad (13)$$

$$g_{pqrs} = \int\int \frac{\phi_p^*(\mathbf{r}_1)\phi_r^*(\mathbf{r}_2)\phi_q(\mathbf{r}_1)\phi_s(\mathbf{r}_2)}{|\mathbf{r}_1 - \mathbf{r}_2|}\, d^3\mathbf{r}_1 d^3\mathbf{r}_2, \quad (14)$$

while the spin adapted one and two electron singlet operators are

$$E_{pq} = \sum_\sigma a^\dagger_{p\sigma}a_{q\sigma}, \quad (15)$$

$$e_{pqrs} = E_{pq}E_{rs} - \delta_{rq}E_{ps}. \quad (16)$$

The fermion operators $a^\dagger_{p\sigma}$ ($a_{p\sigma}$) creates (annihilates) an electron in orbital $p$ with spin $\sigma$. We now consider the photonic operators $\alpha$ and $\beta$ as the symmetric and antisymmetric linear combination of two parent operators $\gamma$ and $\tau$

$$\alpha = \frac{\gamma + \tau}{\sqrt{2}}, \quad (17)$$

$$\beta = \frac{\gamma - \tau}{\sqrt{2}}. \quad (18)$$

Inserting these expression into the Hamiltonian in Eq. 11, we obtain the Hamiltonian in the final form

$$\begin{aligned} H = &\sum_{pq} h_{pq}E_{pq} + \frac{1}{2}\sum_{pqrs}g_{pqrs}e_{pqrs} + \bar{\omega}(\gamma^\dagger\gamma + \tau^\dagger\tau) \\ &- \frac{\lambda}{\sqrt{\bar{\omega}}}\sum_{pq}\left[\mathfrak{Im}\left\{\boldsymbol{\nabla}\cdot\boldsymbol{\epsilon}_\pm e^{i\mathbf{k}\mathbf{r}}\right\}\right]_{pq} E_{pq}(\gamma + \gamma^\dagger) \\ &- \frac{\lambda}{\sqrt{\bar{\omega}}}\sum_{pq}\left[\mathfrak{Re}\left\{\boldsymbol{\nabla}\cdot\boldsymbol{\epsilon}_\pm e^{i\mathbf{k}\mathbf{r}}\right\}\right]_{pq} E_{pq}(\tau + \tau^\dagger), \end{aligned} \quad (19)$$

where we have applied the unitary transformation

$$V = \exp\left(i(\pi\tau^\dagger\tau + \frac{3}{2}\pi\gamma^\dagger\gamma)\right) \quad (20)$$

and used the relation

$$\langle\phi_p|\,(\mathbf{p}\cdot\boldsymbol{\epsilon}_\pm)e^{i\mathbf{k}\mathbf{r}}\,|\phi_q\rangle^* = -\langle\phi_p|\,(\mathbf{p}\cdot\boldsymbol{\epsilon}_\pm^*)e^{-i\mathbf{k}\mathbf{r}}\,|\phi_q\rangle, \quad (21)$$

which specifically holds for real orbitals. The Hamiltonian in Eq. 19 is an effective two-mode Hamiltonian for photons $\gamma$ and $\tau$ with the same frequency $\bar{\omega}$ and field polarization $\boldsymbol{\epsilon}_\pm$. In particular, the velocity gauge interaction between the chiral electromagnetic vacuum and the electrons is mediated by the integrals of the real and imaginary parts of $\boldsymbol{\nabla}\cdot\boldsymbol{\epsilon}_\pm e^{i\mathbf{k}\mathbf{r}}$:

$$\left[\mathfrak{Im}\left\{\boldsymbol{\nabla}\cdot\boldsymbol{\epsilon}_\pm e^{i\mathbf{k}\mathbf{r}}\right\}\right]_{pq}$$
$$= \int \phi_p^*(\mathbf{r})\left[\partial_x \sin(k_z z) \pm \partial_y \cos(k_z z)\right]\phi_q(\mathbf{r})\,d^3\mathbf{r}, \quad (22)$$

$$\left[\mathfrak{Re}\left\{\boldsymbol{\nabla}\cdot\boldsymbol{\epsilon}_\pm e^{i\mathbf{k}\mathbf{r}}\right\}\right]_{pq}$$
$$= \int \phi_p^*(\mathbf{r})\left[\partial_x \cos(k_z z) \mp \partial_y \sin(k_z z)\right]\phi_q(\mathbf{r})\,d^3\mathbf{r}. \quad (23)$$

## 2.2 Strong coupling wave function

The wave function parametrization is derived extending the formalism presented in Ref. [16] for the Hamiltonian in Eq. 19. Specifically, the reference ground state wave function is written as the direct product of a Slater determinant and the electromagnetic vacuum for the $\gamma$ and $\tau$ photons

$$|R\rangle \equiv |HF\rangle \otimes |0_\gamma 0_\tau\rangle. \quad (24)$$

This reference state is transformed with an orbital-specific coherent state transformation

$$U_{\text{SC}} = \exp\left[-\frac{\lambda}{\sqrt{\bar{\omega}^3}}\sum_p \tilde{E}_{pp}\left(\zeta_p\gamma + \xi_p\tau - \text{c.c.}\right)\right]. \quad (25)$$

In Eq. 25, the variational parameters $\{\zeta_p\}$ and $\{\xi_p\}$, associated with photons $\gamma$ and $\tau$ respectively, mix the electronic and photonic degrees of freedom and account for $\omega$-dispersion. This parametrization is obtained from the exact solution in the infinite coupling limit $\lambda \to +\infty$. We point out that the singlet operator $\tilde{E}_{pp}$ in the $U_{\text{SC}}$ transformation should be expressed in the correlated ($\sim$) basis which simultaneously diagonalizes the interaction integrals in Eqs. 22 and 23. However, it is well known that, for a generic multi-mode theory this can only be achieved by employing a complete basis set, such that the canonical position commutators hold, $[\hat{\mathbf{r}}_i, \hat{\mathbf{r}}_j] = 0$, and the interactions commute to allow for their simultaneous diagonalization. In a finite basis set, a correlated basis needs to be chosen and in Sections 3 and 4 we explore two different options, highlighting strengths and weaknesses of each. We also note that Cui *et al.* [56] proposed a different scheme to tackle the multi-mode problem: a diagonal Lang-Firsov coherent state transformation, in which the basis choice is guided by the atomic local orbitals rather than the interaction integrals. More specifically, they applied this transformation to the Hubbard-Holstein Hamiltonian for modeling polaron states in a lattice and to the Pauli-Fierz Hamiltonian in the dipole approximation. The wave function parametrization we adopt is then given by

$$|\psi\rangle = U_{\text{SC}}\exp(\kappa)|R\rangle, \quad (26)$$

where we also employ the $\{\kappa_{ai}\}$ parameters

$$\kappa = \sum_{ai}\kappa_{ai}(E_{ai} - E_{ia}) \quad (27)$$

in the variational optimization of the molecular orbitals as in standard Hartree-Fock theory. Only the occupied-virtual (indices $i$ and $a$) block is considered, since the other rotation blocks are redundant. Finally, we emphasize that adopting the strong coupling parametrization is important for obtaining consistent polaritonic molecular orbitals, especially when orbital features (such as orbital energies) are used in a perturbative description of correlation effects. For a pedagogical derivation of the strong coupling wave function, we refer the reader to the Supplementary Material.

## 2.3 Møller-Plesset perturbation theory

In order to account for electron-electron, chiral electron-photon and photon-photon correlations, we employ the strong coupling Møller-Plesset perturbation theory [17, 56]. Throughout, we first change the quantum picture by applying the $U_{\text{SC}}$ transformation in Eq. 25 to the light-matter Hamiltonian in Eq. 19

$$H_{\text{SC}} = U_{\text{SC}}^\dagger H\, U_{\text{SC}}. \quad (28)$$

Then, we define the zeroth-order Hamiltonian as

$$H_{\text{SC}}^{(0)} = \sum_{pq} F_{pq} E_{pq} + \bar{\omega}(\gamma^\dagger\gamma + \tau^\dagger\tau), \quad (29)$$

where $F_{pq} = \epsilon_p \delta_{pq}$ are the elements of the Fock matrix on the molecular orbital basis, where $\epsilon_p$ is the orbital energy of the orbital $p$. The eigenfunctions for this unperturbed Hamiltonian are the states spanning the polaritonic Hilbert space $\mathcal{H}_{pol} = \mathcal{H}_e \otimes \mathcal{H}_{ph}$

$$H_{\text{SC}}^{(0)} |\mu, n, m\rangle = |\mu, n, m\rangle E_{\mu,n,m}^{(0)}, \quad (30)$$

where the $|\mu\rangle$ states are defined as excitations of the electronic reference $|\text{HF}\rangle$ determinant

$$|\mu\rangle = \tau_\mu |\text{HF}\rangle, \quad (31)$$

with $\tau_\mu$ being an electronic excitation operator. The $n$ and $m$ are the photonic occupation numbers for the $\gamma$ and $\tau$ photons. The eigenvalues in Eq. 30 are given by

$$E_{\mu,n,m}^{(0)} = E_{\text{HF},0_\gamma,0_\tau}^{(0)} + \epsilon_\mu + (n+m)\bar{\omega}, \quad (32)$$

where $E_{\text{HF},0_\gamma,0_\tau}^{(0)}$ is twice the sum of the Hartree-Fock occupied orbital energies and for instance

$$\epsilon_\mu = \epsilon_a - \epsilon_i, \quad \text{for} \quad |\mu\rangle = E_{ai}|\text{HF}\rangle, \quad (33)$$

$$\epsilon_\mu = \epsilon_a - \epsilon_i + \epsilon_b - \epsilon_j, \quad \text{for} \quad |\mu\rangle = E_{bj}E_{ai}|\text{HF}\rangle \quad (34)$$

are the excitation energies for the electronic excited determinants. When we add the first-order correction to the zeroth-order ground state energy, we obtain the mean-field energy obtained after convergence of the variational parameters appearing in the wave function parametrization in Eq. 26. Chiral electron-photon correlation emerges at the second order in the Møller-Plesset perturbative series, together with the electron–electron and photon-photon correlation

$$E_{\text{HF},0_\gamma,0_\tau}^{(2)} = -\sum_\mu \sum_{n,m} \frac{|\langle \text{HF}, 0_\gamma, 0_\tau | H_{\text{SC}} | \mu, n, m \rangle|^2}{E_{\mu,n,m}^{(0)} - E_{\text{HF},0_\gamma,0_\tau}^{(0)}}. \quad (35)$$

In Eq. 35, $\{\mu, n, m\} \neq \{\text{HF}, 0_\gamma, 0_\tau\}$ such that the excitation degree is at least one either in the electronic or photonic subspace. We can further express the second order energy correction in terms of the different possible excitations

$$E_{\text{HF},0_\gamma,0_\tau}^{(2)} = -\sum_{n+m=2}^{\infty} \frac{|E^{nm}|^2}{(n+m)\bar{\omega}}$$
$$-\sum_{n+m=1}^{\infty} \sum_{ai} \frac{2|F_{ai}^{nm}|^2}{\epsilon_a - \epsilon_i + (n+m)\bar{\omega}}$$
$$-\sum_{n+m=0}^{\infty} \sum_{aibj} \frac{g_{aibj}^{nm}(2g_{aibj}^{nm} - g_{ajbi}^{nm})}{\epsilon_a + \epsilon_b - \epsilon_i - \epsilon_j + (n+m)\bar{\omega}}. \quad (36)$$

Specifically, in Eq. 36 we find the contribution from the pure photonic excitations (first line), which are eventually coupled with single and double electronic excitations (in the second and third lines, respectively). While the electronic part does not involve more than double excitations, the photonic part exhibits excitations that extend indefinitely. Therefore, a truncation of the photonic space in some suitable fashion is required as specified in Section 4. For a more detailed description of the terms shown in the Møller-Plesset energy correction, we refer the reader to the Supplementary Material.

## 3 Methods and implementation

In this section, we discuss the implemented method focusing on the variational optimization procedure and related aspects, such as the treatment of photonic redundancies and the choice of the correlated basis. The theory presented in Section 2 has been implemented in a development version of the $e^{\mathcal{T}}$ program [57]. All calculations have been performed using a dual-socket Intel(R) Xeon(R) Platinum 8380 system with 2 TB of memory and 20 cores of CPU.

### 3.1 Parameters variational optimization

The wave function parametrization in Eq. 26 involves two classes of variational parameters: $\{\kappa_{ai}\}$ for the orbital optimization, and $\{\zeta_p\}, \{\xi_p\}$ which account for the chiral photonic dressing of the electrons. All parameters are variationally optimized through a self-consistent field (SCF) procedure, which is iterated until the norm of the global gradient vector

$$\mathbf{E}^{(1)} = \begin{pmatrix} \partial E/\partial \boldsymbol{\kappa} \\ \partial E/\partial \boldsymbol{\zeta} \\ \partial E/\partial \boldsymbol{\xi} \end{pmatrix}_{\boldsymbol{\kappa}=0} \quad (37)$$

is smaller than a chosen convergence threshold. For the polaritonic orbitals, convergence is achieved by solving the Roothaan–Hall equations, where the Fock matrix $\mathbf{F}$ is diagonalized to update the electronic density encoded in the orbital coefficient matrix $\mathbf{C}$. To accelerate convergence, Pulay's direct inversion in the iterative subspace (DIIS) extrapolation is employed [58, 59]. In contrast, a straightforward gradient-based optimization for the photonic parameters would result in slow convergence due to their strong correlation [60]. In Figure 3 we illustrate this correlation with a heat map representation of the $\zeta\xi$-Hessian matrix, computed in the first iteration for the R enantiomer of proline inside a right-handed chiral cavity. The photonic parameters $\zeta$ and $\xi$ exhibit pronounced off-diagonal contributions. For this reason, the optimization of the photonic chiral dressing is performed by computing the Newton steps, $\Delta\zeta$ and $\Delta\xi$, by preconditioning the gradient vector with the inverse of the $\zeta\xi$-Hessian matrix

$$\begin{pmatrix} \Delta\boldsymbol{\zeta} \\ \Delta\boldsymbol{\xi} \end{pmatrix} = - \begin{pmatrix} \frac{\partial^2 E}{\partial\boldsymbol{\zeta}\partial\boldsymbol{\zeta}} & \frac{\partial^2 E}{\partial\boldsymbol{\zeta}\partial\boldsymbol{\xi}} \\ \frac{\partial^2 E}{\partial\boldsymbol{\xi}\partial\boldsymbol{\zeta}} & \frac{\partial^2 E}{\partial\boldsymbol{\xi}\partial\boldsymbol{\xi}} \end{pmatrix}^{-1} \begin{pmatrix} \partial E/\partial\boldsymbol{\zeta} \\ \partial E/\partial\boldsymbol{\xi} \end{pmatrix}. \quad (38)$$

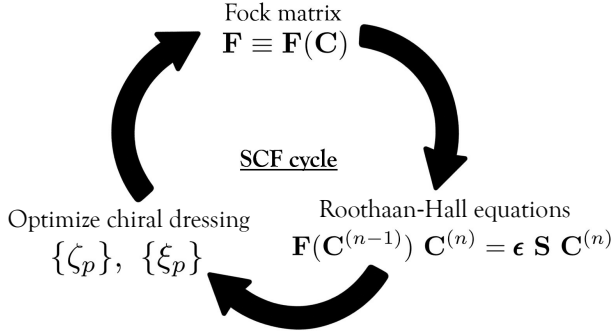

**Fig. 2:** Schematic representation of the self-consistenft field (SCF) cycle for the optimization of the wave function's variational parameters. The dressed electronic density is updated solving the Roothaan-Hall equations for the orbital coefficients $\mathbf{C}$. Finally, before calculation of the updated Fock matrix $\mathbf{F}$, the photonic dressing parameters $\zeta$ and $\xi$ are updated.

For explicit expressions of the gradients and the $\zeta\xi$-Hessian matrix, we refer the reader to the Supplementary Material.

## 3.2 Projection of photonic redundancies

One concern raised in Ref. [17] is the lack of size-intensivity of the polaritonic molecular orbitals, which can arise from bifurcation points introduced by redundancies in the photonic parameter space. Unlike the well known occupied–occupied and virtual–virtual orbital redundancies, these photonic redundancies are more subtle since they are not always associated with a vanishing gradient. Nevertheless, it is essential to project them out during the optimization procedure as they lead to singular values in the $\zeta\xi$-Hessian matrix. This numerically halts the Newton-based convergence of the parameters, which relies on the Hessian matrix inversion. In this work, we compute the $\Delta\zeta$ and $\Delta\xi$ steps in a space orthogonal to the one spanned by the singular eigenvectors $|\nu\rangle$ of the overlap matrix in the correlated basis

$$S_{pq} = \langle\mathrm{HF}|\,\tilde{E}_{pp}\tilde{E}_{qq}\,|\mathrm{HF}\rangle. \quad (39)$$

In doing so, we compute the projector

$$P = I - \sum_\nu |\nu\rangle\langle\nu| \quad (40)$$

and solve Eq. 38 using the projected photonic Hessian matrix and the projected gradient vector. With this procedure, the optimization will correctly handle bifurcation points on the energy surface, preventing convergence to excited states.

## 3.3 Interaction-oriented basis choice

As discussed in Section 2.2, it is necessary to select a correlated basis for the singlet operator in the $U_{\mathrm{SC}}$ transformation of Eq. 25. In a generic multi-mode theory, an interaction-oriented choice can be naturally defined as the coupling-weighted average of the interaction integrals. This approach is exemplified in the case of the dipole-gauge Pauli-Fierz Hamiltonian. There, dipole interaction terms of the form $\mathbf{d}\cdot\boldsymbol{\epsilon}_\alpha$, associated with non-parallel polarizations $\alpha$, do not commute. However, when the polarizations are parallel, a dipole basis can be constructed that is exact in the infinite coupling limit and obtained precisely through such an averaging of the interactions. It follows that, for our effective two-mode Hamiltonian in Eq. 19, we can in principle adopt the basis that diagonalizes the matrix with elements

$$\left[\mathfrak{Im}\left\{\boldsymbol{\nabla}\cdot\boldsymbol{\epsilon}_\pm e^{i\mathbf{kr}}\right\}\right]_{pq} + \left[\mathfrak{Re}\left\{\boldsymbol{\nabla}\cdot\boldsymbol{\epsilon}_\pm e^{i\mathbf{kr}}\right\}\right]_{pq} \quad (41)$$

since both contributions share the same coupling strength $\lambda$. However, this choice introduces two complications. First, although the Hamiltonian itself is origin-invariant (via a phase shift of the momentum operator $\boldsymbol{\nabla}$), the use of such a basis leads to origin-dependent results. This dependence arises from the explicit exponential factors that involve position $\mathbf{r}$. Illustrative examples of this effect are provided in the Supplementary Material. Second, performing two calculations with opposite handedness of the polarization vector would require different bases, which prevents a direct comparison when evaluating the discriminating power. To circumvent these issues, we also employ a basis that diagonalizes the sum of the interactions within the dipole approximation. In this case, we use a linear real polarization vector oriented along the same direction as

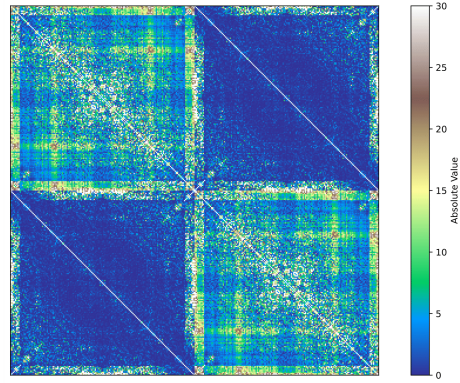

$$\mathbf{E}_{ph}^{(2)} = \begin{pmatrix} \frac{\partial^2 E}{\partial \boldsymbol{\zeta} \partial \boldsymbol{\zeta}} & \vdots & \frac{\partial^2 E}{\partial \boldsymbol{\zeta} \partial \boldsymbol{\xi}} \\ \cdots & \vdots & \cdots \\ \frac{\partial^2 E}{\partial \boldsymbol{\xi} \partial \boldsymbol{\zeta}} & \vdots & \frac{\partial^2 E}{\partial \boldsymbol{\xi} \partial \boldsymbol{\xi}} \end{pmatrix} =$$

**Fig. 3:** Heat map representation of the photonic $\zeta\xi$-Hessian matrix computed at the first SCF cycle for the S enantiomer of a proline molecule confined within a right-handed circularly polarized (RHCP) chiral cavity. The frequency and the light-matter coupling are set respectively to $\omega = 2.72$ eV and $\lambda = 0.005$ a.u. The calculation has been performed using the aug-cc-pVDZ basis set [61, 62]. The photonic $\zeta$ and $\xi$ parameters appear to be strongly correlated requiring a second order optimization algorithm for their convergence.

the original chiral polarization, with matrix elements given by

$$\int \phi_p^*(\mathbf{r})(\partial_x + \partial_y)\phi_q(\mathbf{r})\, d^3\mathbf{r}. \quad (42)$$

One remark is in order regarding this basis choice. The dipole approximation is applied only in the basis selection, while the Hamiltonian itself retains the full spatial structure of the electromagnetic field. This ensures that the enantioselective effects are still accounted for. Lastly, we mention that for both choices of basis in Eqs. 41 and 42, evaluation of integrals in a complex basis is necessary, effectively doubling the memory requirements of the calculation. This drawback can be mitigated through a complex Cholesky decomposition of the two electron integrals in the selected correlated basis [63–66]. The optimal solution for the correlated basis would be to not choose any option *a priori*, but to variationally optimize it.

## 4 Results and Discussions

In this section, we evaluate the performance of the proposed method in describing the enantioselective discriminating power of chiral electromagnetic vacuum fluctuations when coupled to a chiral molecular system. For the calculation of the second order Møller-Plesset energy corrections, accounting for electron–electron, chiral electron-photon and photon-photon correlation, the infinite summations in the photonic space in Eq. 36 are truncated when the contributions are smaller than a threshold set to $10^{-12}$ a.u.

In Figure 4, we show the energy difference between the R and S enantiomers ($\Delta$RS) of alanine confined within a left-handed circularly polarized (LHCP) chiral cavity, evaluated across different cavity frequencies $\omega$. Specifically, we report both the mean-field and the Møller–Plesset corrected energies, using the two basis choices introduced in Eqs. 42 and 41: panels (a) and (b). The molecule is centered at the origin and for all calculations we employ a cc-pVDZ basis set [61, 62]. We compare these results with those shown in Figure 5, which displays the $\omega$-dispersion of the same light–matter system and basis set computed using the minimal coupling QED-CCSD method [44]. In this approach, the coupled cluster calculations are based on Hartree–Fock orbitals as the reference wave function, while electron–photon correlation effects are incorporated through the minimal coupling Hamiltonian of Eq. 11 and the cluster operator truncated to include up to single photonic excitations. All calculations were carried out using a light–matter coupling strength of $\lambda = 0.005$ a.u., which corresponds to a quantization volume of approximately 75 nm$^3$. This volume is notably larger than those currently achievable in Fabry-Pérot cavities, which can be decreased to 1 nm$^3$. In Figure 5, the S enantiomer is consistently more stabilized within the LHCP chiral cavity across the entire inspected frequency range (up to 30 eV), with a minimum in the enantiomeric discrimination power around 4 eV. This behavior is reproduced by the second order Møller–Plesset energy on the right of panel (a) in Figure 4, where the minimum shifts to a slightly higher cavity frequency, 5 eV, while maintaining the same order of magnitude, 0.1 $\mu$eV. On the left of panel (a) in Figure 4, the pure mean-field result obtained with the same correlation basis fails to reproduce the correct trend. It instead predicts the R enantiomer to be more stable, showing a maximum rather than a minimum and a larger magnitude of the discrimination power. This highlights the importance of the correlation effects, which are described perturbatively beyond the mean-field level. This demonstrates that cavity–frequency correlation, included at the mean-field level, is insufficient to describe the complexity of the many-body system composed by electrons and chiral vacuum photons. In panel (b) of Figure 4, on the other hand, we observe very different trends in the enantiomeric discrimination power as a function of the cavity frequency $\omega$. Here, the basis is chosen such that it diagonalizes the sum of the interactions retaining the full shape of the field. With this choice, the mean-field and second order Møller–Plesset energies exhibit again opposite behaviors of $\Delta$RS, which is on the order of 0.1 meV at low frequencies and monotonically decreasing in the order of magnitude and eventually reversing sign around $\omega = 30$ eV for the Møller-Plesset correction. Note that, calculating the response of the opposite enantiomer under a circular polarization is equivalent to calculating for the same enantiomer under the opposite helicity of the polarization vector. Upon further analysis of this behavior, we argue that the trends displayed in panel (b) of Figure 4 arise primarily from the use of different bases in the two calculations needed for calculating $\Delta$RS, rather than retaining the Fourier exponentials in Eq. 41. In fact, we obtain the same results shown in panel (a) of Figure 4 when the multipolar form of the electromagnetic fields is retained, while employing a real linear polarization for the basis choice. This behavior is probably due to the position of the molecular system, see the Supplementary Material for an example where, on the other hand, the effect is apparent due to the shift of the origin. We suggest that allowing the basis to dynamically adapt across different frequencies, through the dispersion relation $|\mathbf{k}| = \omega/c$, can provide flexibility to improve convergence for molecular systems located at different positions relative to the origin. This consideration is particularly relevant for extended molecular systems, such as ensembles of many molecules in solution, where defining a unique center of mass becomes ambiguous and collective light–matter effects emerge [67–72]. In future work, the incorporation of cavity boundary conditions by introducing additional parameters into the origin-dependent basis may provide a promising direction to explore.

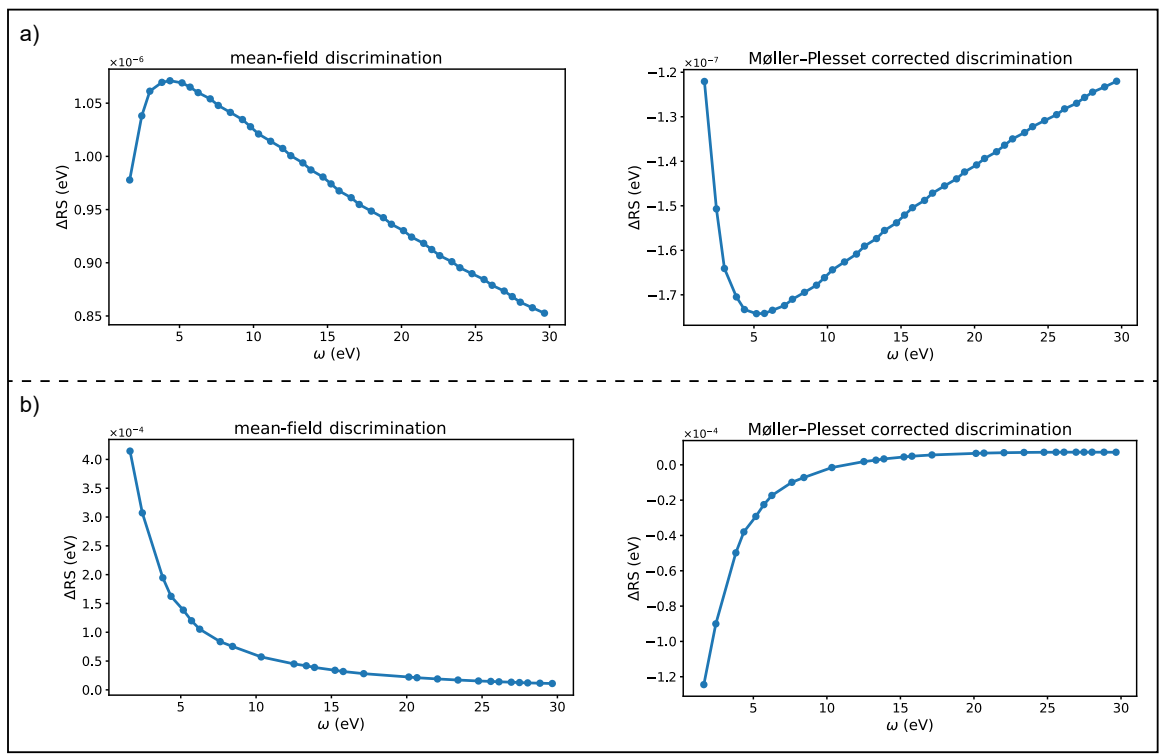

**Fig. 4:** Mean-field and second order Møller–Plesset corrected enantiomeric energy discrimination $\Delta$RS as a function of the cavity frequency $\omega$ for an alanine molecule confined within a left-handed circularly polarized (LHCP) chiral cavity. The light–matter coupling strength is fixed at $\lambda = 0.005$ a.u. Panel (a) shows the results obtained using the basis that diagonalizes the sum of interaction integrals while retaining the full spatial structure and chirality of the fields (Eq. 41), whereas panel (b) reports the results obtained with the basis constructed within the dipole approximation and a real linear polarization (Eq. 42). The molecular geometry is placed at the origin of the light–matter system, and all calculations are performed with the cc-pVDZ basis set.

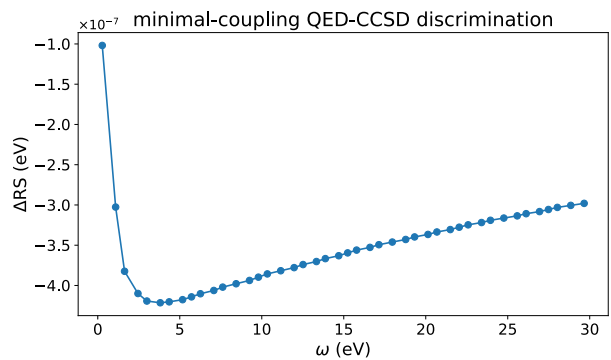

**Fig. 5:** Minimal coupling QED-CCSD enantiomeric energy discrimination $\Delta$RS as a function of the cavity frequency $\omega$ for an alanine molecule confined within a left-handed circularly polarized (LHCP) chiral cavity. The light–matter coupling strength is fixed at $\lambda = 0.005$ a.u. and the cc-pVDZ basis set is employed.

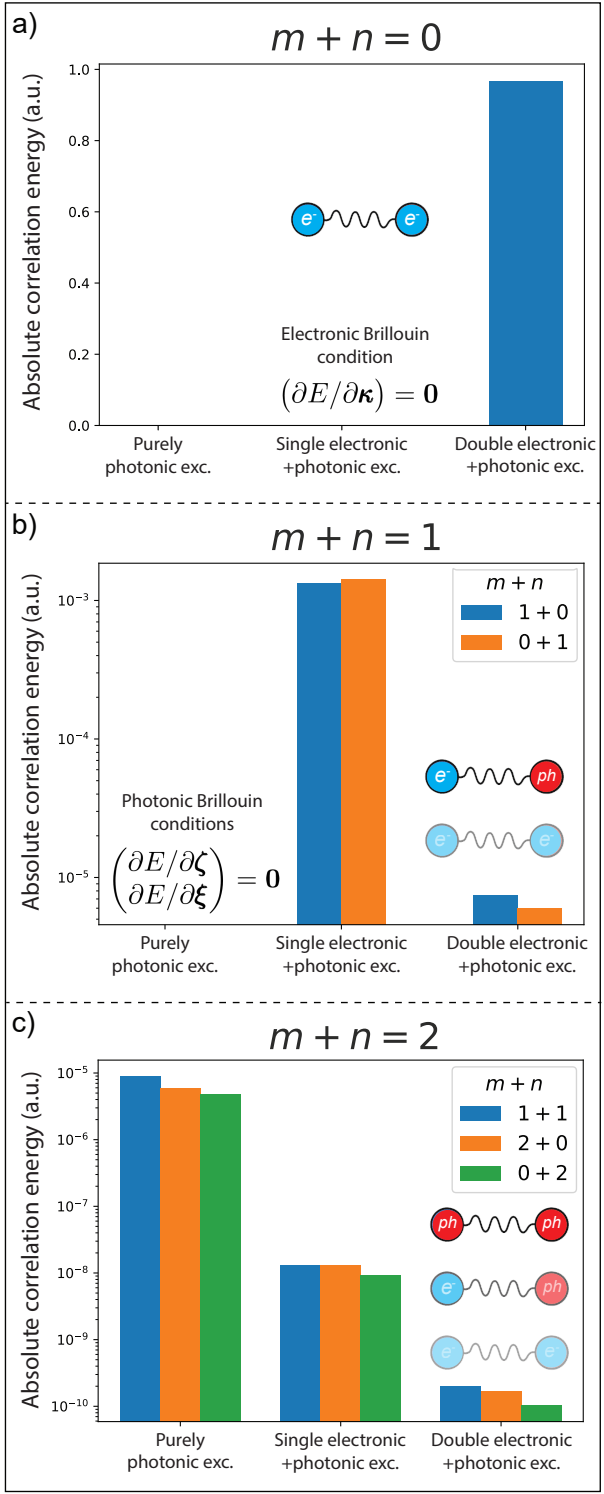

**Fig. 6:** Absolute value contributions to the correlation energy captured with strong coupling Møller-Plesset second order perturbation theory for different degrees of the global photonic excitation $m + n$: panels (a), (b) and (c). The results refer to an S enantiomer of an alanine molecule confined in a LHCP chiral cavity with cavity frequency $\omega = 1.67$ eV, light-matter coupling $\lambda = 0.005$ a.u. and the cc-pVDZ basis set. Chiral electron-photon correlation energy is captured through mixed excitations.

In Figure 6, we show the various contributions to the correlation energy perturbatively given by Eq. 36 for the second order of the strong coupling Møller–Plesset perturbation theory. The light-matter system is composed of an S enantiomer of alanine confined in a chiral LHCP cavity with cavity frequency $\omega = 1.67$ eV and light-matter coupling $\lambda = 0.005$ a.u.. The results have been obtained by employing the basis that diagonalizes the integrals in Eq. 42, using the cc-pVDZ basis set. Specifically, we show in panels (a), (b) and (c) the contributions for different $m + n$ degrees of the total photonic excitation, and classify them into three different categories: purely photonic excitations, photonic excitations coupled with single electronic excitations and photonic excitations coupled with double electronic excitations. Contributions with $m + n \geq 3$ are smaller than $10^{-12}$ a.u. and are therefore neglected. The largest contribution, on the order of unity, arises from the purely electron–electron correlation associated with double electronic excitations at $m + n = 0$ in panel (a). For higher photonic excitations given in panels (b) and (c), the chiral electron-photon correlation contributes together with further electronic correlation, with a magnitude that systematically decreases with the total electron–photon excitation degree. For $m + n = 2$ in panel (c), we also observe purely photon–photon correlation contributions with a magnitude on the order of $10^{-5}$ a.u. In the case of mixed elec-

tron–photon excitations, all three types of correlation contribute simultaneously, making the analysis more intricate since their effects are non-additive and they enter nonlinearly in the second order correction. Moreover, a direct comparison with the correlation energy at the coupled cluster level is challenging for two main reasons. First, the currently available minimal coupling QED-CCSD implementation only account for single photonic excitations in the cluster operator. Nevertheless, contributions from higher degrees of excitation do arise indirectly through the unlinked terms inherent in coupled cluster theory. The last feature is advantageous in regards to coupled cluster theory, but represents also the second reason why the comparison with perturbation theory is not straightforward. Comparing with an extended coupled cluster parametrization capable of describing linked double photonic excitations could provide further insights into the complex interplay between electrons and chiral vacuum photons.

Finally, we investigate how the introduction of multiple chiral centers around a central chiral molecule influences the enantiomeric discrimination power. In Ref. [44] it has been shown that an achiral solvent, such as water, can enhance the discrimination through solvent-induced effects, where the first solvation shells reorganize into a chiral arrangement. However, the role of a chiral solvent has not yet been explored. As a simple model, we considered the systems shown in Figure 7, consisting of a methyloxirane molecule interacting with two tetrahydrofuran (THF) molecules, which are fluorinated to introduce additional stereogenic centers. We performed calculations placing these molecular systems inside a RHCP chiral cavity with cavity frequency $\omega = 13.6$ eV and light–matter coupling strength $\lambda = 0.001$ a.u. corresponding to a quantization volume of approximately 600 nm$^3$. In Table 1, the resulting enantiomeric discriminations for the overall system, $\Delta RS$, are compared also considering the discrimination for the isolated methyloxirane and fluoro-THF molecules under the same cavity conditions. For all the calculations we used a cc-pVDZ basis set, and for the mean-field and Møller-Plesset corrected energies we employed the basis that diagonalizes the integrals in Eq. 42. We first observe that the perturbation theory results gives the correct sign of the discrimination power for the isolated molecules, in agreement with coupled cluster theory, and therefore provide a reliable reference. When comparing mean-field and perturbation theory, we find that the addition of two non-chiral THF molecules enhances the absolute value of the discrimination power. The sign predicted by perturbation theory is considered trustworthy, as it matches both the isolated methyloxirane result and the coupled cluster reference. Introducing stereogenic centers into the THF molecules further amplifies the discrimination, but also reverses its sign. This inversion originates from the opposite discrimination tendencies of fluoro-THF and methyloxirane. However, this trend is not shown by the coupled cluster calculations, where the discrimination power remains very small in magnitude. The coupled cluster result after introduction of two stereogenic centers is smaller than our convergence threshold of $10^{-12}$ a.u. and therefore is not reported. This discrepancy may be attributed to the absence of linked double photonic excitations in the current implementation of minimal-coupling QED-CCSD. Further investigation, particularly on solvent effects, will be necessary to clarify their impact on the discrimination power.

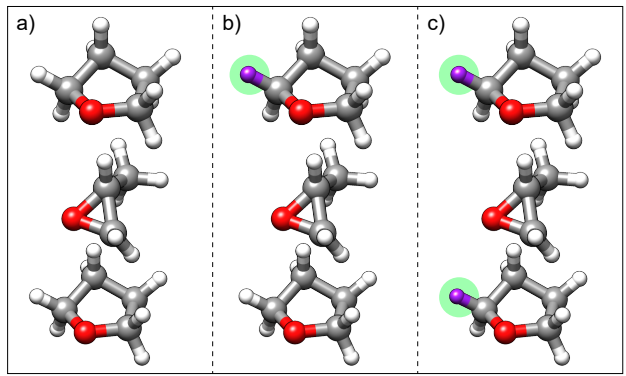

**Fig. 7:** Methyloxirane molecule interacting with two tetrahydrofuran (THF) molecules in panel (a), which are fluorinated to introduce additional stereogenic centers in panels (b) and (c).

**Tab. 1:** Enantiomeric discrimination power for methyloxirane, fluoro-THF and the systems showed in Figure 7. All the molecular systems are placed in a RHCP chiral cavity with cavity frequency $\omega = 13.6$ eV and light–matter coupling strength $\lambda = 0.001$ a.u. We employed the cc-pVDZ basis set for all calculations.

| molecular system | $\Delta RS$ (eV) mean-field | $\Delta RS$ (eV) Møller-Plesset | $\Delta RS$ (eV) coupled cluster |
|---|---|---|---|
| methyloxirane | $-4.50 \cdot 10^{-8}$ | $-8.6 \cdot 10^{-9}$ | $-1.23 \cdot 10^{-8}$ |
| fluoro-THF | $5.54 \cdot 10^{-8}$ | $1.72 \cdot 10^{-8}$ | $1.08 \cdot 10^{-8}$ |
| panel (a) Fig. 7 | $5.30 \cdot 10^{-6}$ | $-9.98 \cdot 10^{-7}$ | $-1.05 \cdot 10^{-8}$ |
| panel (b) Fig. 7 | $8.82 \cdot 10^{-6}$ | $2.27 \cdot 10^{-6}$ | $-8.5 \cdot 10^{-9}$ |
| panel (c) Fig. 7 | $1.15 \cdot 10^{-5}$ | $4.48 \cdot 10^{-6}$ | – |

# 5 Conclusions

In this work, we develop and implement a theoretical framework for modeling the ground state strong coupling between electrons and chiral vacuum photons inside a circularly polarized chiral cavity. For the energy modeling, we employ the full minimal coupling Hamiltonian effectively considering two counter-propagating chiral photonic modes and reabsorbing the important field-squared diamagnetic term into a cavity-dressed frequency. The reference wave function parametrization is derived in order to be exact in the infinite coupling limit extending the strong coupling QED Hartree-Fock theory to a multi-mode multipolar Hamiltonian. The variational parameters are self-consistently optimized using a Newton-based algorithm for the photonic space, accurately accounting and projecting out redundancies in order to prevent singularities in the Hessian matrix. The choice of the correlated basis emerges as an important nontrivial delicate matter due to origin-dependence and computation of the enantiomeric discrimination power. Specifically, we propose and discuss different ways to choose an interaction-oriented basis. The chiral electron-photon correlation effects responsible for the enantioselective interaction are described going beyond the mean-field approximation using strong coupling Møller-Plesset perturbation theory up to second order.

Comparing the QED mean-field, Møller–Plesset perturbation theory and coupled cluster models, our results show that the mean-field approximation, although able to capture cavity frequency $\omega$-dispersion, is insufficient to capture the correct qualitative behavior of the enantiomeric discrimination. Instead, the inclusion of chiral electron-photon correlation in the perturbation theory is essential to reproduce the stabilization trends observed with coupled cluster theory. The comparison between different basis choices further reveals a pronounced sensitivity, underscoring the importance of using the same basis for calculating the energy difference of two enantiomers within the same chiral cavity. Nonetheless a frequency-adaptive basis may be useful, particularly for extended molecular systems where the collective regime emerges. Perturbative analysis of the correlation contributions shows that pure electron–electron interactions dominate, while mixed electron–photon excitations are responsible for the emergence of enantioselectivity from higher-order photonic excitations and providing systematically smaller but still non-negligible contributions. Although direct comparison with minimal coupling QED-CCSD is complicated due to the differences in how the excitation manifolds are accessed, our study highlights the complementarity of perturbative and coupled-cluster approaches and points to the need for extended coupled-cluster parametrizations that explicitly include multiple linked photonic excitations.

Overall, this work shows that capturing the enantioselective power of chiral light requires going beyond mean-field theory and establishes a foundation for more accurate and general descriptions of molecular chirality in quantum cavity environments. Looking ahead, promising directions include explicitly incorporating chiral cavity boundary conditions through tailored basis choices, enhancing discrimination power with strong magnetic fields [73, 74], and pumping real chiral photons into the cavity to induce chirality in otherwise insensitive systems [26]. Further developments may also focus on response theory [75–77], enabling access to chiral polaritonic states with Rabi splitting [24], and electronic circular dichroism (ECD) spectra within chiral optical devices.

**Research funding:** YEM, RRR and HK acknowledge funding from the European Research Council (ERC) under the European Union's Horizon 2020 Research and Innovation Programme (grant agreement No. 101020016).

**Acknowledgment:** We acknowledge insightful discussions with Federico Rossi and Matteo Castagnola, and we appreciate their comments on the manuscript.

**Author contributions:** YEM, RRR and HK conceived the project. YEM carried out the theoretical modeling, the implementation of the method, and performed the numerical simulations. AB implemented the complex phase interaction integrals [78]. YEM and RRR prepared the initial draft of the manuscript. HK supervised the project. All authors discussed the results and revised the manuscript.

**Data availability statement:** The $e^{\mathcal{T}}$ outputs are available at [79]. The code can be made available upon reasonable request to the authors.

**Conflict of interest:** Authors state no conflict of interest.

**Supplementary Material:** This article contains supplementary material, including the derivation of the strong coupling wave function parametrization for the polaritonic ground state, detailed equations for the parameter gradients and the photonic Hessian matrix, explicit terms for the strong coupling second order Møller–Plesset energy correction, origin-dependent discrimination power results obtained through basis choices that retain the spatial shape of the electromagnetic field.

# Supplementary Material to:

# Unveiling chiral electron–photon correlation effects in circularly polarized optical devices

Yassir El Moutaoukal,[†] Rosario R. Riso,[†] Andrea Bianchi,[‡] and Henrik Koch[*,†]

[†]*Department of Chemistry, Norwegian University of Science and Technology, 7491 Trondheim, Norway*

[‡]*Scuola Normale Superiore, Piazza dei Cavalieri, 7, 56126 Pisa PI, Italy*

E-mail: henrik.koch@ntnu.no

# Table of Contents

# S1. Strong coupling wave function parametrization

Our objective is to parametrize a wave function suitable for the ground-state description of the effective two-mode light–matter Hamiltonian

$$H = \sum_{pq} h_{pq} E_{pq} + \frac{1}{2} \sum_{pqrs} g_{pqrs} e_{pqrs} + \bar{\omega}(\gamma^\dagger \gamma + \tau^\dagger \tau)$$
$$- \frac{\lambda}{\sqrt{\bar{\omega}}} \sum_{pq} \left[ \mathfrak{Im} \left\{ \boldsymbol{\nabla} \cdot \boldsymbol{\epsilon}_\pm e^{i\mathbf{kr}} \right\} \right]_{pq} E_{pq} (\gamma + \gamma^\dagger) \quad \text{(SE1)}$$
$$- \frac{\lambda}{\sqrt{\bar{\omega}}} \sum_{pq} \left[ \mathfrak{Re} \left\{ \boldsymbol{\nabla} \cdot \boldsymbol{\epsilon}_\pm e^{i\mathbf{kr}} \right\} \right]_{pq} E_{pq} (\tau + \tau^\dagger)$$

able to model the interaction between a molecular system with the electromagnetic vacuum fluctuations manifesting inside a chiral cavity.[1] Following the strategy presented in Ref. 2, we consider the infinite coupling limit ($\lambda \to +\infty$) of the Hamiltonian.
In this limit, the Hamiltonian simplifies to

$$H_\infty = \bar{\omega}\left(\gamma^\dagger - \frac{\lambda}{\sqrt{\bar{\omega}^3}} \sum_{pq} \left[\mathfrak{Im}\left\{(\boldsymbol{\nabla} \cdot \boldsymbol{\epsilon})e^{i\mathbf{kr}}\right\}\right]_{pq} E_{pq}\right)\left(\gamma - \frac{\lambda}{\sqrt{\bar{\omega}^3}} \sum_{pq} \left[\mathfrak{Im}\left\{(\boldsymbol{\nabla} \cdot \boldsymbol{\epsilon})e^{i\mathbf{kr}}\right\}\right]_{pq} E_{pq}\right)$$
$$+ \bar{\omega}\left(\tau^\dagger - \frac{\lambda}{\sqrt{\bar{\omega}^3}} \sum_{pq} \left[\mathfrak{Re}\left\{(\boldsymbol{\nabla} \cdot \boldsymbol{\epsilon})e^{i\mathbf{kr}}\right\}\right]_{pq} E_{pq}\right)\left(\tau - \frac{\lambda}{\sqrt{\bar{\omega}^3}} \sum_{pq} \left[\mathfrak{Re}\left\{(\boldsymbol{\nabla} \cdot \boldsymbol{\epsilon})e^{i\mathbf{kr}}\right\}\right]_{pq} E_{pq}\right)$$
$$- \frac{\lambda^2}{\bar{\omega}^2} \sum_{pqrs} \left[\mathfrak{Im}\left\{(\boldsymbol{\nabla} \cdot \boldsymbol{\epsilon})e^{i\mathbf{kr}}\right\}\right]_{pq} \left[\mathfrak{Im}\left\{(\boldsymbol{\nabla} \cdot \boldsymbol{\epsilon})e^{i\mathbf{kr}}\right\}\right]_{rs} (e_{pqrs} + E_{ps}\delta_{rq})$$
$$- \frac{\lambda^2}{\bar{\omega}^2} \sum_{pqrs} \left[\mathfrak{Re}\left\{(\boldsymbol{\nabla} \cdot \boldsymbol{\epsilon})e^{i\mathbf{kr}}\right\}\right]_{pq} \left[\mathfrak{Re}\left\{(\boldsymbol{\nabla} \cdot \boldsymbol{\epsilon})e^{i\mathbf{kr}}\right\}\right]_{rs} (e_{pqrs} + E_{ps}\delta_{rq}),$$
$$\text{(SE2)}$$

where the free-field contribution $H_{ph} = \bar{\omega}(\gamma^\dagger \gamma + \tau^\dagger \tau)$ is retained in order to ensure that the Hamiltonian remains bounded from below. The last two terms in Eq. (SE1) play the role of velocity-gauge multipolar analogues of the dipole self-energy (DSE) terms appearing in the length-gauge Pauli–Fierz Hamiltonian under the dipole approximation. The exact eigenfunctions of $H_\infty$ have electronic parts that are Slater determinants expressed in the basis that simultaneously diagonalizes the real and imaginary parts of the integrals $\left[(\boldsymbol{\nabla}\cdot\boldsymbol{\epsilon})e^{i\mathbf{kr}}\right]_{pq}$. In practice, such a simultaneous diagonalization is possible only in the limit of a complete set, such that the canonical position commutators hold, $[\hat{\mathbf{r}}_i, \hat{\mathbf{r}}_j] = 0$, and the interaction terms commute. In this hypothetical complete correlated basis ($\sim$), the Hamiltonian in the infinite coupling limit takes the form

$$H_\infty = \bar{\omega}\left(\gamma^\dagger - \frac{\lambda}{\sqrt{\bar{\omega}^3}}\sum_p \left[\widetilde{\mathfrak{Im}}\left\{(\boldsymbol{\nabla}\cdot\boldsymbol{\epsilon})e^{i\mathbf{kr}}\right\}\right]_{pp}\tilde{E}_{pp}\right)\left(\gamma - \frac{\lambda}{\sqrt{\bar{\omega}^3}}\sum_p \left[\widetilde{\mathfrak{Im}}\left\{(\boldsymbol{\nabla}\cdot\boldsymbol{\epsilon})e^{i\mathbf{kr}}\right\}\right]_{pp}\tilde{E}_{pp}\right)$$
$$+ \bar{\omega}\left(\tau^\dagger - \frac{\lambda}{\sqrt{\bar{\omega}^3}}\sum_p \left[\widetilde{\mathfrak{Re}}\left\{(\boldsymbol{\nabla}\cdot\boldsymbol{\epsilon})e^{i\mathbf{kr}}\right\}\right]_{pp}\tilde{E}_{pp}\right)\left(\tau - \frac{\lambda}{\sqrt{\bar{\omega}^3}}\sum_p \left[\widetilde{\mathfrak{Re}}\left\{(\boldsymbol{\nabla}\cdot\boldsymbol{\epsilon})e^{i\mathbf{kr}}\right\}\right]_{pp}\tilde{E}_{pp}\right)$$
$$- \frac{\lambda^2}{\bar{\omega}^2}\sum_{pq}\left[\widetilde{\mathfrak{Im}}\left\{(\boldsymbol{\nabla}\cdot\boldsymbol{\epsilon})e^{i\mathbf{kr}}\right\}\right]_{pp}\left[\widetilde{\mathfrak{Im}}\left\{(\boldsymbol{\nabla}\cdot\boldsymbol{\epsilon})e^{i\mathbf{kr}}\right\}\right]_{qq}(\tilde{e}_{ppqq} + \tilde{E}_{pq}\delta_{pq})$$
$$- \frac{\lambda^2}{\bar{\omega}^2}\sum_{pq}\left[\widetilde{\mathfrak{Re}}\left\{(\boldsymbol{\nabla}\cdot\boldsymbol{\epsilon})e^{i\mathbf{kr}}\right\}\right]_{pp}\left[\widetilde{\mathfrak{Re}}\left\{(\boldsymbol{\nabla}\cdot\boldsymbol{\epsilon})e^{i\mathbf{kr}}\right\}\right]_{qq}(\tilde{e}_{ppqq} + \tilde{E}_{pq}\delta_{pq}),$$
(SE3)

where, for compactness, we use the notation $\widetilde{\mathfrak{Re}}$ and $\widetilde{\mathfrak{Im}}$ to indicate the interaction integrals evaluated in the correlated basis. We can change the quantum picture by introducing a transformation able to reduce the limit Hamiltonian to the purely free-field and multipolar self-interaction parts

$$H_\infty \xrightarrow{U_\infty^\dagger H_\infty U_\infty} H_\infty = \bar{\omega}(\gamma^\dagger\gamma + \tau^\dagger\tau)$$
$$+ \frac{\lambda^2}{\bar{\omega}^2}\sum_{pq}\left[\widetilde{\mathfrak{Im}}\left\{(\mathbf{p}\cdot\boldsymbol{\epsilon})e^{i\mathbf{kr}}\right\}\right]_{pp}\left[\widetilde{\mathfrak{Im}}\left\{(\mathbf{p}\cdot\boldsymbol{\epsilon})e^{i\mathbf{kr}}\right\}\right]_{qq}(\tilde{e}_{ppqq} + \tilde{E}_{pq}\delta_{pq}) \quad \text{(SE4)}$$
$$+ \frac{\lambda^2}{\bar{\omega}^2}\sum_{pq}\left[\widetilde{\mathfrak{Re}}\left\{(\mathbf{p}\cdot\boldsymbol{\epsilon})e^{i\mathbf{kr}}\right\}\right]_{pp}\left[\widetilde{\mathfrak{Re}}\left\{(\mathbf{p}\cdot\boldsymbol{\epsilon})e^{i\mathbf{kr}}\right\}\right]_{qq}(\tilde{e}_{ppqq} + \tilde{E}_{pq}\delta_{pq}).$$

The gauge transformation able to perform such a shifting of the bosonic operators is

$$U_\infty = \exp\left(-\frac{\lambda}{\sqrt{\bar{\omega}^3}}\sum_p E_{pp}\left(\left[\widetilde{\mathfrak{Im}}\left\{(\boldsymbol{\nabla}\cdot\boldsymbol{\epsilon})e^{i\mathbf{kr}}\right\}\right]_{pp}(\gamma - \gamma^\dagger) + \left[\widetilde{\mathfrak{Re}}\left\{(\boldsymbol{\nabla}\cdot\boldsymbol{\epsilon})e^{i\mathbf{kr}}\right\}\right]_{pp}(\tau - \tau^\dagger)\right)\right).$$
(SE5)

We now relax the infinite coupling limit to the case of a finite light–matter interaction strength. In this regime, the untransformed Hamiltonian naturally recovers the purely electronic contribution $H_{el} = \sum_{pq} \tilde{h}_{pq} \tilde{E}_{pq} + \frac{1}{2}\sum_{pqrs} \tilde{g}_{pqrs} \tilde{e}_{pqrs}$. On the other hand, in the transformation in eq. (SE5) we substitute the diagonal elements of the real and imaginary parts of the integrals $(\boldsymbol{\nabla}\cdot\boldsymbol{\epsilon})e^{i\mathbf{k}\mathbf{r}}$ with two sets of orbital-specific parameters: $\{\xi_p\}$, the $\xi$-parameters, and $\{\zeta_p\}$, the $\zeta$-parameters,

$$U_{\text{SC}} = \exp\left(-\frac{\lambda}{\sqrt{\tilde{\omega}^3}} \sum_p \tilde{E}_{pp} \left(\zeta_p(\gamma - \gamma^\dagger) + \xi_p(\tau - \tau^\dagger)\right)\right). \quad \text{(SE6)}$$

This transformation mixes the electronic and photonic degrees of freedom by mean of the introduced $\xi$ and $\zeta$ parameters, which have to be variationally optimized in order to account for the chiral photonic dressing of the electrons. Note that, for realistic computations, the use of a complete basis set is clearly unfeasible. For this reason we should consider the projection of the Hamiltonian to a finite basis set where the interaction integrals integrals $\left[(\boldsymbol{\nabla}\cdot\boldsymbol{\epsilon})e^{i\mathbf{k}\mathbf{r}}\right]_{pq}$ are not diagonal. We point out that the shape of SC-transformation in eq. (SE6) can remain in its diagonal form even in a finite basis set because the coherent-state parameters $\xi$ and $\zeta$ can always be redirected to a diagonal form. The $\sim$ symbol refers now to a generic correlated basis of our choice. Then, the chiral SC wave function parametrization reads

$$|\psi\rangle = \exp\left(-\frac{\lambda}{\sqrt{\tilde{\omega}^3}} \sum_p \tilde{E}_{pp} \left(\zeta_p(\gamma - \gamma^\dagger) + \xi_p(\tau - \tau^\dagger)\right)\right) e^{\kappa} |\text{HF}\rangle \otimes |0_\gamma 0_\tau\rangle, \quad \text{(SE7)}$$

where $|\text{HF}\rangle$ and $|0_\gamma 0_\tau\rangle$ are respectively the Hartree-Fock Slater determinant for the electrons and the photonic vacuum. For sake of simpler notation, in later equations we define the reference ground-state $|\text{R}\rangle \equiv |\text{HF}\rangle \otimes |0_\gamma 0_\tau\rangle$. Moreover, in our wave function parametrization we also make use of the exponential $\kappa$-parametrization for optimizing the polaritonic molecular orbitals:

$$\kappa = \sum_{ai} \kappa_{ai}(E_{ai} - E_{ia}). \quad \text{(SE8)}$$

# S2. Gradients, $\zeta\xi$-Hessian and SCF optimization

We can rewrite the effective two-mode light-matter Hamiltonian in eq. (SE1) as

$$H = \sum_{pq} h_{pq} E_{pq} + \frac{1}{2} \sum_{pqrs} g_{pqrs} e_{pqrs} + \bar{\omega} \sum_{\rho \in \{\gamma,\tau\}} b_\rho^\dagger b_\rho \\ - \sum_{\rho \in \{\gamma,\tau\}} \sum_{pq} g_{\rho,pq} E_{pq}(b_\rho + b_\rho^\dagger),$$
(SE9)

where the interaction integrals $g_{\rho,pq}$ are

$$g_{\gamma,pq} = \frac{\lambda}{\sqrt{\bar{\omega}}} \left[ \mathfrak{Im}\left\{ (\boldsymbol{\nabla} \cdot \boldsymbol{\epsilon}) e^{i\mathbf{kr}} \right\} \right]_{pq},$$
(SE10)

$$g_{\tau,pq} = \frac{\lambda}{\sqrt{\bar{\omega}}} \left[ \mathfrak{Re}\left\{ (\boldsymbol{\nabla} \cdot \boldsymbol{\epsilon}) e^{i\mathbf{kr}} \right\} \right]_{pq}.$$
(SE11)

On the other hand, the wave function parametrization can be rewritten as

$$|\psi\rangle = \prod_c \exp\left(-\frac{1}{\bar{\omega}} \sum_p \eta_p^\rho \tilde{E}_{pp}(b_\rho - b_\rho^\dagger)\right) e^\kappa |\mathrm{R}\rangle,$$
(SE12)

where the parameters $\eta_p^\rho$ are

$$\eta_p^\gamma = \frac{\lambda}{\sqrt{\bar{\omega}}} \zeta_p,$$
(SE13)

$$\eta_p^\tau = \frac{\lambda}{\sqrt{\bar{\omega}}} \xi_p.$$
(SE14)

In total, the wave function parametrization is then composed by two classes of parameters: $\{\kappa_{ai}\}$ and $\{\eta_p^\rho\}$, where $\rho \in \{\gamma,\tau\}$.

Before proceeding with the derivation of the gradients and photonic Hessian equations, it is useful to show the effect of the effective two-mode SC-transformation

$$U_{\mathrm{SC}} = \prod_{\rho \in \{\gamma,\tau\}} \exp\left(-\frac{1}{\bar{\omega}} \sum_p \eta_p^\rho \tilde{E}_{pp}(b_\rho - b_\rho^\dagger)\right)$$
(SE15)

to the Hamiltonian in eq. (SE9):

$$\begin{aligned} H_{\text{SC}} &= U_{\text{SC}}^\dagger\, H\, U_{\text{SC}} \\ &= \sum_{pq} \tilde{h}_{pq}^{\text{SC}} \pi_{pq} \tilde{E}_{pq} + \frac{1}{2} \sum_{pqrs} \tilde{g}_{pqrs}^{\text{SC}} \pi_{pqrs} \tilde{e}_{pqrs} + \bar{\omega} \sum_{\rho \in \{\gamma,\tau\}} b_\rho^\dagger b_\rho \\ &\quad - \sum_{\rho \in \{\gamma,\tau\}} \sum_{pq} g_{\rho,pq}^\eta \tilde{E}_{pq} (b_\rho^\dagger \pi_{pq} + \pi_{pq} b_\rho), \end{aligned} \qquad \text{(SE16)}$$

where the redefined one and two electron SC-integrals are

$$\tilde{h}_{pq}^{\text{SC}} = \tilde{h}_{pq} + \frac{1}{\bar{\omega}} \sum_{\rho \in \{\gamma,\tau\}} \left( \sum_r (g_{\rho,pr}^\eta g_{\rho,rq}^\eta - g_{\rho,pr} g_{\rho,rq}) \right), \qquad \text{(SE17)}$$

$$\tilde{g}_{pqrs}^{\text{SC}} = \tilde{g}_{pqrs} + \frac{2}{\bar{\omega}} \sum_{\rho \in \{\gamma,\tau\}} (g_{\rho,pq}^\eta g_{\rho,rs}^\eta - g_{\rho,pq} g_{\rho,rs}), \qquad \text{(SE18)}$$

the $\eta$-shifted interaction integrals $g_{\rho,pq}^\eta$ read

$$g_{\rho,pq}^\eta = g_{\rho,pq} - \eta_p^\rho \delta_{pq} \qquad \text{(SE19)}$$

and finally the bosonic $\pi_{pq}^\gamma$ and $\pi_{pqrs}^\gamma$ operators are defined as

$$\pi_{pq} = \prod_{\sigma \in \{\gamma,\tau\}} \exp\left( \frac{1}{\bar{\omega}} (\eta_p^\sigma - \eta_q^\sigma)(b_\sigma - b_\sigma^\dagger) \right), \qquad \text{(SE20)}$$

$$\pi_{pqrs} = \prod_{\sigma \in \{\gamma,\tau\}} \exp\left( \frac{1}{\bar{\omega}} (\eta_p^\sigma - \eta_q^\sigma + \eta_r^\sigma - \eta_s^\sigma)(b_\sigma - b_\sigma^\dagger) \right). \qquad \text{(SE21)}$$

In the following, we calculate the gradients at $\boldsymbol{\kappa} = \mathbf{0}$ by considering that the canonical basis is updated at each iteration. Seemingly, we calculate the gradients also at the updated $\eta$-parameters coming from the previous iteration step.

The energy gradient for the chiral SC method is composed by two classes: $\kappa$ and $\eta$

$$\mathbf{E}^{(1)} = \begin{pmatrix} \partial E/\partial \boldsymbol{\kappa} \\ \partial E/\partial \boldsymbol{\eta} \end{pmatrix}. \tag{SE22}$$

Due to the high correlation between the $\eta$-parameters, we use the $\eta\eta$-Hessian

$$\mathbf{E}^{(2)}_{\eta\eta} = \left( \partial^2 E/\partial \boldsymbol{\eta}^2 \right) \tag{SE23}$$

for preconditioning the $\eta$-step in a Newton-based algorithm presented in Ref. 3.

## S2.1 Orbital optimization: $\kappa$-gradient

The $\kappa$-gradient elements are defined

$$\frac{\partial E}{\partial \kappa_{ai}} = \langle \mathrm{R} | [H_{\mathrm{SC}}, E_{ai}^-] | \mathrm{R} \rangle = 2F_{ai}, \tag{SE24}$$

where $F_{ai}$ are virtual-occupied Fock matrix elements in the canonical basis and obtained from the rotation of the correlated basis Fock matrix

$$F_{pq} = \sum_{rs} V_{pr} \tilde{F}_{rs} V_{qs}. \tag{SE25}$$

More specifically, $V$ is the orthonormal matrix connecting the two basis and

$$\tilde{F}_{pq} = \tilde{h}^{\Pi}_{pq} + \frac{1}{2} \sum_{rs} (2\tilde{g}^{\Pi}_{pqrs} - \tilde{g}^{\Pi}_{psrq}) \tilde{D}_{rs}. \tag{SE26}$$

The $\Pi$-redefined one and two electron integrals read as

$$\tilde{h}^{\Pi}_{pq} = \tilde{h}^{\mathrm{SC}}_{pq} \prod_{\sigma \in \{\gamma, \tau\}} \exp\left( -\frac{1}{2\bar{\omega}^2} (\eta^{\sigma}_p - \eta^{\sigma}_q)^2 \right), \tag{SE27}$$

$$\tilde{g}^{\Pi}_{pqrs} = \tilde{g}^{\text{SC}}_{pqrs} \prod_{\sigma \in \{\gamma, \tau\}} \exp\left(-\frac{1}{2\bar{\omega}^2}(\eta^\sigma_p - \eta^\sigma_q + \eta^\sigma_r - \eta^\sigma_s)^2\right), \tag{SE28}$$

where the gaussian factors derive from the vacuum average of the bosonic operators in eqs. (SE20) and (SE21).

For the polaritonic orbitals, convergence is obtained by iteratively solving the Roothaan–Hall equations, where diagonalization of the Fock matrix $\mathbf{F}$ provides updated orbital coefficients collected in the matrix $\mathbf{C}$, which in turn define the electronic density. To improve the rate of convergence, Pulay's direct inversion in the iterative subspace (DIIS) method is applied.

## S2.2 Chiral photon dressing optimization: $\eta$-gradient and $\eta\eta$-Hessian

The $\eta$-gradient vector is composed by two modes contributions

$$\left(\frac{\partial E}{\partial \boldsymbol{\eta}}\right) = \begin{pmatrix} \frac{\partial E}{\partial \boldsymbol{\eta}^\gamma} \\ \frac{\partial E}{\partial \boldsymbol{\eta}^\tau} \end{pmatrix}. \tag{SE29}$$

Each $\eta^\rho$-gradient element is defined as

$$\begin{aligned}
\frac{\partial E}{\partial \eta^\rho_m} &= \frac{1}{\bar{\omega}} \langle \text{R}| [\tilde{E}_{pp}(b_\rho - b^\dagger_\rho), H_{\text{SC}}] |\text{R}\rangle \\
&= \frac{2}{\bar{\omega}^2} \sum_q (\tilde{h}^\Pi_{mq}\tilde{D}_{mq} - \tilde{h}^\Pi_{qm}\tilde{D}_{qm})\Delta^\rho_{mq} \\
&\quad + \frac{2}{\bar{\omega}^2} \sum_{qrs} (\tilde{g}^\Pi_{mqrs}\tilde{d}_{mqrs} - \tilde{g}^\Pi_{qmrs}\tilde{d}_{qmrs})\Delta^\rho_{mqrs} \\
&\quad - \frac{1}{\bar{\omega}} \sum_q (g^\Pi_{\rho,mq}\tilde{D}_{mq} + g^\Pi_{\rho,qm}\tilde{D}_{qm}) - \frac{2}{\bar{\omega}} \sum_{rs} g^\Pi_{\rho,rs}\tilde{d}_{mmrs},
\end{aligned} \tag{SE30}$$

where the $\Pi$-redefined interaction integrals read

$$g^\Pi_{\rho,pq} = g^\eta_{\rho,pq} \prod_{\sigma \in \{\gamma, \tau\}} \exp\left(-\frac{1}{2\bar{\omega}^2}(\eta^\sigma_p - \eta^\sigma_q)^2\right), \tag{SE31}$$

and the $\Delta$-factors are defined as negative the differences between the $\eta^\rho$-parameters

$$\Delta^\rho_{pq} = -(\eta^\rho_p - \eta^\rho_q), \tag{SE32}$$

$$\Delta^\rho_{pqrs} = -(\eta^\rho_p - \eta^\rho_q + \eta^\rho_r - \eta^\rho_s). \tag{SE33}$$

The $\eta\eta$-Hessian has a block-diagonal structure where the off-diagonal terms represent the coupling between the $\eta$-parameters belonging to different modes. The $\eta\eta$-Hessian elements

$$\begin{aligned}
\frac{\partial^2 E}{\partial \eta^\mu_m \partial \eta^\nu_n} &= \frac{1}{\bar{\omega}^2} \langle \mathrm{R} | \left[ \tilde{E}_{nn}(b_\nu - b^\dagger_\nu), \left[ \tilde{E}_{mm}(b_\mu - b^\dagger_\mu), H_{\mathrm{SC}} \right] \right] | \mathrm{R} \rangle \\
&= \frac{2}{\bar{\omega}^2} \tilde{h}^\Pi_{mn} \Omega^{\mu\nu}_{mn} \tilde{D}_{mn} - \delta_{mn} \frac{2}{\bar{\omega}^2} \left( \sum_q \tilde{h}^\Pi_{nq} \Omega^{\mu\nu}_{nq} \tilde{D}_{nq} + \sum_{qrs} \tilde{g}^\Pi_{nqrs} \Omega^{\mu\nu}_{nqrs} \tilde{d}_{nqrs} \right) \\
&\quad + \frac{2}{\bar{\omega}^2} \sum_{rs} (\tilde{g}^\Pi_{mnrs} \Omega^{\mu\nu}_{mnrs} \tilde{d}_{mnrs} - \tilde{g}^\Pi_{mrns} \Omega^{\mu\nu}_{mrns} \tilde{d}_{mrns} + \tilde{g}^\Pi_{mrsn} \Omega^{\mu\nu}_{mrsn} \tilde{d}_{mrsn}) \\
&\quad + \delta_{\mu\nu} \frac{2}{\bar{\omega}} (\tilde{d}_{mmnn} + \delta_{mn} \tilde{D}_{mm}) - \frac{4}{\bar{\omega}^3} \sum_q \left( g^\Pi_{\mu,nq} \Delta^\nu_{nq} \tilde{d}_{mmnq} + g^\Pi_{\nu,mq} \Delta^\mu_{mq} \tilde{d}_{nnmq} \right) \\
&\quad - \delta_{mn} \frac{2}{\bar{\omega}^3} \sum_q \left( g^\Pi_{\mu,nq} \Delta^\nu_{nq} + g^\Pi_{\nu,nq} \Delta^\mu_{nq} \right) \tilde{D}_{nq} + \frac{2}{\bar{\omega}^3} \left( g^\Pi_{\mu,mn} \Delta^\nu_{mn} - g^\Pi_{\nu,mn} \Delta^\mu_{mn} \right) \tilde{D}_{mn}
\end{aligned} \tag{SE34}$$

where the $\Omega$-factors are defined from the $\Delta$-factors

$$\Omega^{\mu\nu}_{pq} = \delta_{\mu\nu} - \frac{1}{\bar{\omega}^2} \Delta^\mu_{pq} \Delta^\nu_{pq}, \tag{SE35}$$

$$\Omega^{\mu\nu}_{pqrs} = \delta_{\mu\nu} - \frac{1}{\bar{\omega}^2} \Delta^\mu_{pqrs} \Delta^\nu_{pqrs}. \tag{SE36}$$

Direct gradient optimization of the photonic dressing parameters is hindered by strong correlations. We instead compute Newton steps using the inverse $\zeta\xi$-Hessian, while projecting out photonic redundancies that cause singular values and stall convergence. The steps are evaluated in the subspace orthogonal to the singular eigenvectors of the correlated basis overlap matrix, using the projected Hessian and gradient.

# S3. Second-order Møller-Plesset energy correction

We compute the second-order energy correction with Møller-Plesset perturbation theory presented in Refs. 4 and 5 in order to capture electron-electron and chiral electron-photon correlations effects. The energy correction to the mean-field energy reads

$$
\begin{aligned}
E^{(2)}_{\text{HF},0_\gamma,0_\tau} = &- \sum_{n+m=2}^{\infty} \frac{|E^{nm}|^2}{(n+m)\bar{\omega}} \\
&- \sum_{n+m=1}^{\infty} \sum_{ai} \frac{2|F^{nm}_{ai}|^2}{\epsilon_a - \epsilon_i + (n+m)\bar{\omega}} \\
&- \sum_{n+m=0}^{\infty} \sum_{aibj} \frac{g^{nm}_{aibj}(2g^{nm}_{aibj} - g^{nm}_{ajbi})}{\epsilon_a + \epsilon_b - \epsilon_i - \epsilon_j + (n+m)\bar{\omega}},
\end{aligned}
\quad (\text{SE37})
$$

where we have contributions from purely photonic excitations in first row, which can be eventually coupled with single or double electronic excitations respectively showing in second and third rows. Starting from purely photonic excitations, the contributions with $n+m=1$ are null due to the Brillouin conditions in the photonic blocks

$$
\langle R | \left[ H_{\text{SC}}, \tilde{E}_{pp}(b_\rho - b_\rho^\dagger) \right] | R \rangle = 0, \quad \text{for } \rho \in \{\gamma, \tau\}. \quad (\text{SE38})
$$

The energy term showing in the numerator is defined

$$
E^{nm} = 2 \sum_i h^{nm}_{ii} + \sum_{ij} (2 g^{nm}_{iijj} - g^{nm}_{ijji}), \quad (\text{SE39})
$$

where the the $nm$-th one and two electron integrals in the canonical basis are

$$
\begin{aligned}
h^{nm}_{pq} = &\frac{1}{\sqrt{n!m!}} \sum_{rs} V_{pr} \, \tilde{h}^{\Pi}_{rs} \left( \frac{1}{\bar{\omega}} (\eta^\gamma_r - \eta^\gamma_s) \right)^n \left( \frac{1}{\bar{\omega}} (\eta^\tau_r - \eta^\tau_s) \right)^m V_{qs} \\
&+ (1-\delta_{n0}) \sqrt{\frac{n}{(n-1)!m!}} \sum_{rs} V_{pr} \, \tilde{g}^{\Pi}_{\gamma,rs} \left( \frac{1}{\bar{\omega}} (\eta^\gamma_r - \eta^\gamma_s) \right)^{n-1} \left( \frac{1}{\bar{\omega}} (\eta^\tau_r - \eta^\tau_s) \right)^m V_{qs} \\
&+ (1-\delta_{m0}) \sqrt{\frac{m}{n!(m-1)!}} \sum_{rs} V_{pr} \, \tilde{g}^{\Pi}_{\tau,rs} \left( \frac{1}{\bar{\omega}} (\eta^\gamma_r - \eta^\gamma_s) \right)^n \left( \frac{1}{\bar{\omega}} (\eta^\tau_r - \eta^\tau_s) \right)^{m-1} V_{qs},
\end{aligned}
\quad (\text{SE40})
$$

$$g^{nm}_{pqrs} = \frac{1}{\sqrt{n!m!}} \sum_{tuvz} V_{pt} V_{rv} \, \tilde{g}^{\text{II}}_{tuvz} \left(\frac{1}{\bar{\omega}}(\eta^\gamma_t - \eta^\gamma_u + \eta^\gamma_v - \eta^\gamma_z)\right)^n \left(\frac{1}{\bar{\omega}}(\eta^\tau_t - \eta^\tau_u + \eta^\tau_v - \eta^\tau_z)\right)^m V_{qu} V_{sz}.$$
(SE41)

These integrals has been obtained using the displacement Franck-Condon factors:

$$\langle l|e^{-\alpha(b-b^\dagger)}|o\rangle = \begin{cases} \sqrt{\frac{o!}{l!}} \, \alpha^{l-o} e^{-\alpha^2/2} L_o^{l-o}(\alpha^2), & l \geq o \\ \sqrt{\frac{l!}{o!}} \, (-\alpha)^{o-l} e^{-\alpha^2/2} L_o^{o-l}(\alpha^2), & l < o \end{cases}$$
(SE42)

where $L^q_p$ is the Laguerre $q$-th order polynomial of degree $p$. Moving to single electronic excitations, the contributions with $n + m = 0$ is null due to the Brillouin conditions in the electronic block

$$\langle \text{R}| \left[H_{\text{SC}}, E^-_{ai}\right] |\text{R}\rangle = 0.$$
(SE43)

Using Slater-Condon rules, $nm$-th Fock matrix elements showing in the numerator read

$$F^{nm}_{pq} = \sum_{rs} V_{pr} \tilde{F}^{nm}_{rs} V_{qs}$$
(SE44)

and $\tilde{F}^{nm}_{pq}$ are in the correlated basis

$$\tilde{F}^{nm}_{pq} = \tilde{h}^{nm}_{pq} + \sum_{rs}(2\tilde{g}^{nm}_{pqrs} - \tilde{g}^{nm}_{psrq})\tilde{D}_{rs}.$$
(SE45)

Lastly, using Slater-Condon rules, we have contribution from double electronic excitations eventually coupled with photonic excitations that resemble the purely electronic correction of standard Møller-Plesset perturbation theory.

# S4. Origin dependent basis choice

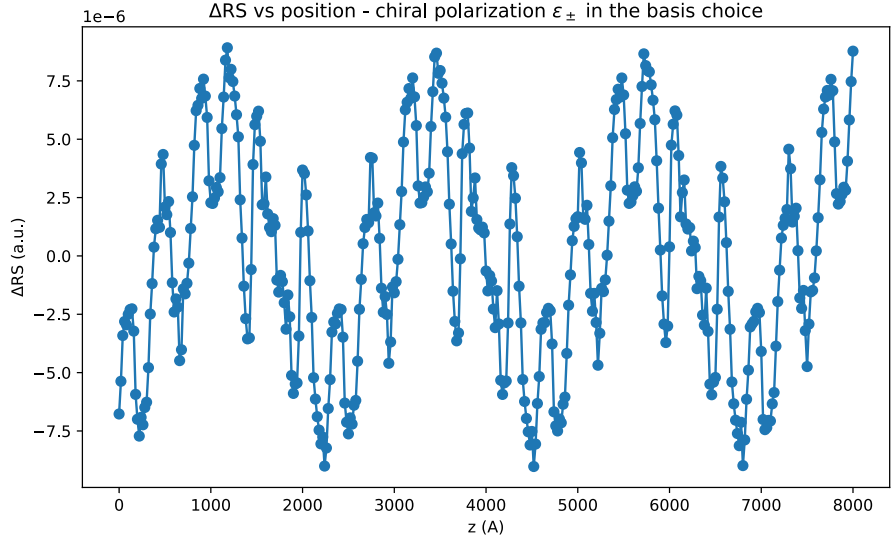

Figure 1: Enantiomeric discrimination power for a methyloxirane molecule confined in a LHCP chiral cavity as a function of its position along the wave vector $\hat{z}$ direction. The cavity-frequency is set to $\omega = 2.72$ eV, while the light-matter coupling strength is $\lambda = 0005$ a.u. The calculations employed the 6-31G basis set and the basis is chosen such to diagonalize the sum of the interaction integrals in eqs. (SE10) and (SE11)

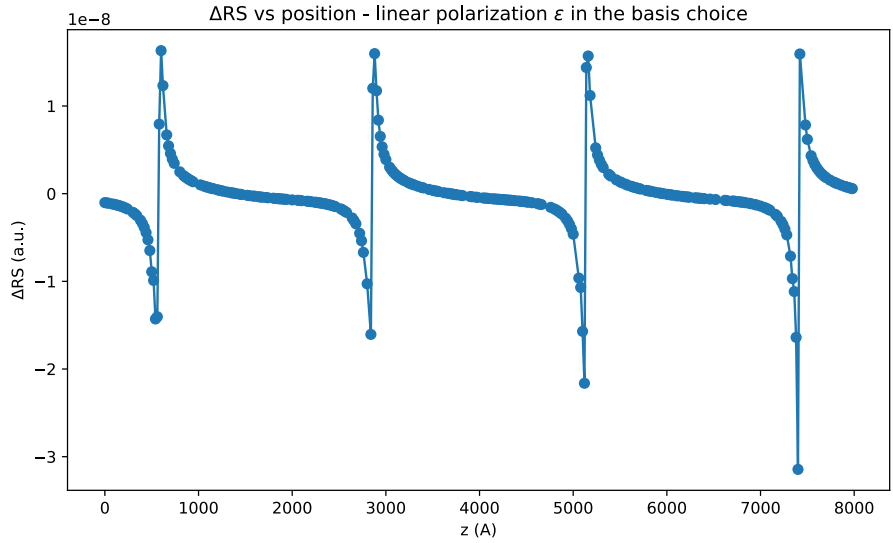

Figure 2: Enantiomeric discrimination power for a methyloxirane molecule confined in a LHCP chiral cavity as a function of its position along the wave vector $\hat{z}$ direction. The cavity-frequency is set to $\omega = 2.72$ eV, while the light-matter coupling strength is $\lambda = 0005$ a.u. The calculations employed the 6-31G basis set and the basis is chosen such to diagonalize the sum of the interaction integrals in eqs. (SE10) and (SE11), but considering a real linear polarization $\epsilon$ along the same $\frac{(\hat{x}+\hat{y})}{\sqrt{2}}$ direction.